# Understanding the spreading patterns of mobile phone viruses

Pu Wang[1,2], Marta C. González[1], César A. Hidalgo[1,2,3] & Albert-László Barabási [1,4]

*[1]Center for Complex Network Research, Department of Physics,*
*Biology and Computer Science, Northeastern University, Boston, MA 02115.*
*[2]Center for Complex Network Research and Department of Physics,*
*University of Notre Dame, Notre Dame, IN 46556.*
*[3]Center for International Development, Kennedy School of Government,*
*Harvard University, Cambridge, MA 02139.*
*[4]Department of Medicine, Harvard Medical School, and Center for Cancer Systems Biology,*
*Dana Farber Cancer Institute, Boston, MA 02115.*

**We model the mobility of mobile phone users to study the fundamental spreading patterns characterizing a mobile virus outbreak. We find that while Bluetooth viruses can reach all susceptible handsets with time, they spread slowly due to human mobility, offering ample opportunities to deploy antiviral software. In contrast, viruses utilizing multimedia messaging services could infect all users in hours, but currently a phase transition on the underlying call graph limits them to only a small fraction of the susceptible users. These results explain the lack of a major mobile virus breakout so far and predict that once a mobile operating system's market share reaches the phase transition point, viruses will pose a serious threat to mobile communications.**

Lacking a standardized operating system, traditional cellphones have been relatively immune to viruses. Smartphones, however, can share programs and data with each other, representing a fertile ground for virus writers (*1-4*). Indeed, since 2004 more than 420 smartphone viruses have been identified (*2*, *3*), the newer ones having reached a state of sophistication that took computer viruses about two decades to arrive at (*2*). While smartphones currently represent less than 5% of the mobile market, given their reported high annual growth rate (*4*) they are poised to become the dominant communication device in the near future, raising the possibility of virus breakouts that could overshadow the disruption caused by traditional computer viruses (*5*).

The spread of mobile viruses is aided by two dominant communication protocols. First, a Bluetooth virus can infect all Bluetooth-activated phones within a distance from 10 to 30m, resulting in a spatially localized spreading pattern similar to the one observed in the case of influenza (*3*, *6*, *7*), SARS (*8*, *9*) and other contact-based diseases (*10*) (Fig. 1A). Second, an MMS virus can send a copy of itself to all mobile phones whose numbers are found in the infected phone's address book, a long range spreading pattern previously exploited only by computer viruses (*11*, *12*). Thus to quantitatively study the spreading dynamics of mobile viruses we need to simultaneously model the location (*13*), the mobility (*14-17*) and the communication patterns (*18-21*) of mobile phone users. To achieve this we use as input anonymized billing record of a mobile phone provider, providing the calling patterns and the coordinates of the closest mobile phone tower each time mobile subscribers use their phone (but not the coordinate of individual users).

The methods we used to track the spreading of a potential Bluetooth and MMS virus are described in the Supporting Online Material (SOM). Briefly, once a phone becomes infected with an MMS virus, after a $\tau$=2min time it sends a copy of itself to each mobile phone number found in the handset's phone book, approximated with the list of numbers the handset's user communicated with during a month long observational period. A Bluetooth virus can infect only mobile phones within a distance $r$=10m. To track this process, we assign to each user an hourly location that is consistent with its travel patterns (*13*) and follow the infection dynamics within each mobile tower area using the SI model (*22*). That is, we consider that an infected user (*I*) infects a susceptible user (*S*), so that the number of infected users evolves in time as $dI/dt=\beta SI/N$, where the effective infection rate is $\beta=\mu<k>$ with $\mu$=1 and the average number of contacts is $<k>=\rho A=NA/A_{tower}$, where $A=\pi r^2$ represents the Bluetooth communication area and $\rho=N/A_{\text{tower}}$ is the population density inside a tower's service area. Once an infected user moves in the vicinity of a new tower, it will serve as a source of a Bluetooth infection in its new location.

A cell phone virus can infect only the phones with the operating system (OS) it was designed for (*2*, *3*), making the market share *m* of an OS an important free parameter in our study. Note that the current market share of various smartphone OSs vary widely, from as little as 2.6% for Palm OS, to 64.3% of Symbian. Given that smartphones together represent less than 5% of all phones, the overall market share of these operating systems among all mobile phones is in the range of $m \simeq 0.0013$ for Palm OS and $m \simeq 0.032$ for Symbian, numbers that are expected to dramatically increase as smartphones replace traditional phones. To maintain the generality of our results, we treat *m* as a free parameter, finding that the spreading of both Bluetooth and MMS viruses is highly sensitive to the market share of the susceptible handsets (Fig. 2, A and B). Our simulations indicate that given sufficient time, a Bluetooth virus can reach all susceptible handsets, as user mobility guarantees that sooner or later each susceptible handset will find itself in the vicinity of an infected handset. The spreading rate strongly depends, however, on the handset's market share. For example if the handset's market share is $m=0.01$, it takes several months for a Bluetooth virus to reach all susceptible handsets. In contrast, for $m=0.30$ the Bluetooth virus could infect 85% of susceptible handsets in a few hours and 99.8% in less than a week (Fig. 2A).

The most striking difference between Bluetooth and MMS viruses comes in the timescales their spread requires. Indeed, given that it takes approximately two minutes for a typical MMS virus to copy itself on a new handset (*23*), an MMS epidemic reaches saturation in a few hours in contrast with a few days it requires the Bluetooth virus to infect all susceptible handsets (Fig. 2B). Thus while there is plenty of time to deploy an antiviral software for a Bluetooth virus before it could reach a large fraction of users, it is largely impossible to achieve the same for MMS viruses, given their explosive spread. The good news is that an MMS virus can reach only a small *m*-dependent fraction of users with a susceptible handset, as indicated by the saturation of the infection curves in Fig. 2B. The origin of this saturation is the fragmentation of the underlying call network. Indeed, in Fig. 1B we show a subset of the real call network and assume for illustration that the handsets can have only two OSs ($OS_1$ and $OS_2$) with market shares $m_1$=0.75 and $m_2$=0.25. While the underlying call network itself is fully connected, the call graph of the users that share the same handset is fragmented into many islands (Fig. 1C). For $m_1$=0.75 we observe a giant component (the largest connected cluster, Fig. 1C) of size $G_m$=0.80, meaning that it contains 80% of the users with the $OS_1$ handset, the rest of the $OS_1$ users being scattered in small isolated clusters. In contrast, for the $OS_2$ handsets the giant component is tiny ($G_m$=0.06). If an MMS virus is released from a single handset, it can only reach the handsets in the cluster where the original handset is located, telling us that an MMS virus can infect at most a $G_m$ fraction of all susceptible handsets, which is 80% for $OS_1$ and 6% for $OS_2$ in the example of Fig. 1.

We find that the handset based fragmentation of the call graph (Fig. 1, B and C) is governed by a percolation phase transition at the market share $m_c$=0.095 (Fig. 2C) (*24*). That is, for $m<m_c$ the user base is fragmented into many small isolated islands, making a major MMS virus viral outbreak impossible. In contrast, for handsets with $m>m_c$ there is a giant component, allowing the MMS virus to reach all handsets that are part of it. The value of $m_c$ and $G_m$ for $m>m_c$ can be calculated using the generating function formalism (*25*), requiring as input only the network's degree distribution $P(k)$. Using $P(k)$ charactering our user base, we find a reasonable agreement between the analytical predictions and the direct measurements of the saturation value of the MMS virus spreading in the mobile phone dataset (Fig. 2C), the small systematic deviation being rooted in the fact that the generating function formalism ignores the correlations in the call graph's structure. The significance of Fig. 2C comes in its ability to explain why we have not observed a significant MMS outbreak so far: currently the market share of the largest OS is less than $m \simeq 0.03$, well under the predicted percolation transition point $m_c \simeq 0.095$ (*26, 27*). For a more detailed discussion on the factors affecting $m$ and $m_c$ see the SOM.

The differences between MMS and Bluetooth viruses have a strong impact on their spreading dynamics as well. To see this we denote with $T(q,m)$ the latency time, representing the average time necessary for a virus affecting a handset with market share $m$ to reach a $q$ fraction of all susceptible handsets. For a Bluetooth virus $T(q,m)$ is finite for any $q$ and $m$ combination, given that with time the virus can reach all susceptible users (Fig. 2A). We find, however, that the latency time is highly sensitive to $m$, a dependency well approximated by $T(q,m) \sim m^{-0.6}$ (Fig. 2D, R-square>0.99, see SOM for statistical analysis), implying that the smaller a handset's market share, the longer it will take for a virus to reach a $q$ fraction of susceptible users. The observed divergence at $m=0$ indicates that for handsets with small market share the spreading process is exceptionally slow, as it takes a very long time for an infected user to come in contact with another user with a similar handset.

Once again, the behavior of MMS viruses is qualitatively different: we find that $T(q,m)$ diverges not at $m=0$ but at a finite $m_q^*$ value (Fig. 2E), meaning that for handsets with market share $m<m_q^*$ the virus is unable to reach a $q$ fraction of users. Indeed, an MMS virus can reach at most a $G_m$ fraction of eligible handsets (Fig. 2C), implying that $G_m$ acts as a critical point for the dynamical spreading process and $T(q>G_m, m)=\infty$. To characterize the observed singularity, we note that the maximum amount of time it takes an MMS virus to invade the giant component should be determined by the length of the longest minimal path $L_{max}$ characterizing the susceptible giant cluster (*28, 29*). As Fig. 2F shows, we find that both $L_{max}$ and the average minimal path length $L_{ave}$ diverge as $(m-m^*)^{-\alpha}$ with $\alpha \approx 0.2$ (R-square>0.97, see SOM), a singularity that potentially drives the observed divergence of $T(q,m)$ in the vicinity of $m_q^*$ given by the equation $q=G\,m_q^*$. A more detailed measurement indicates, however, a systematic $q$-dependence of $\alpha(q)$ in $T(q,m) \sim (m-m_q^*)^{-\alpha(q)}$ in the vicinity of the critical point (see Fig. 2E and the statistical analysis in SOM), hinting that there are factors beyond $L_{max}$ that contribute to the divergence of $T(q,m)$.

In Figure 3 we follow the spread of an MMS and Bluetooth infection starting from the same user, illustrating that Bluetooth and MMS viruses differ in their spatial spreading patterns as well: a Bluetooth virus follows a wave like pattern, infecting predominantly users in the vicinity of the virus's release point, while an MMS virus follows a more delocalized pattern, given that the users' address book often contains phone numbers of far away individuals. To quantify the observed differences we measured the average distance between the cell phone tower where the first infected user is located and the location of towers servicing the newly infected users. A null model in which the virus always diffuses to the non-infected towers bordering the already infected towers, thus following a classical two-dimensional diffusion process, is used as a reference. As Fig. 3B indicates, the typical source-infection distances observed in the local model are significantly smaller than the distances recorded for either Bluetooth or MMS viruses, indicating the impact of a few long-distance travellers that incubate outbreaks in distant cells (*13*) in the Bluetooth spreading process. The average distance is the highest for MMS viruses, underlying the delocalized pattern charactering its spread. Fig. 3B also shows that the dependence of $<D>$ on $N$ is mainly a function of the spreading technology and appears to be independent $m$.

Bluetooth and MMS viruses have their relative limitations: while the spread of a Bluetooth virus is rather slow due to human mobility, an MMS virus can reach only a small fraction of users due to the fragmentation of the call graph. Both limitations are avoided by hybrid viruses that can simultaneously use both Bluetooth and MMS connections to spread, the first of many such viruses being the "CommWarrior" released in 2005 (*2*, *3*). We find, however, that the spreading dynamics of a hybrid virus also displays a complex market share dependence (Fig. 4, A and B), resulting from a nontrivial superposition of the Bluetooth and the MMS spreading modes. For example for $m$=0.15, when there is a giant component aiding the MMS spreading mode (Fig. 2C), the early stage of the spreading process is dominated by the rapid invasion of the MMS cluster. Subsequently, the Bluetooth mechanism allows the virus to invade the rest of the independent MMS clusters as well. For $m$=0.01, however, there is no MMS giant component, thus the spreading is dominated entirely by the Bluetooth capability, resulting in a significantly slower spreading pattern (note the different horizontal axes in Fig. 4, A and B).

The relative role of the Bluetooth and the MMS spreading patterns for hybrid viruses is illustrated in Fig. 4, C and D, which show the latency time $T(q,m)$ for $q$=0.15 and $q$=0.65. We find that for high $m$ the MMS mechanism dominates the hybrid virus's spreading pattern. As $m$ decreases below $m_q^*$ given by $q=G\,m_q^*$ (Fig. 2E) the giant component becomes smaller than $q$, so $T(q,m)$ for MMS diverges (green curve) and the Bluetooth mechanism starts dominating the spreading rate of hybrid virus. Therefore for small $m$ the latency time of the hybrid virus converges to the latency time of a Bluetooth virus. We find, however, that the phase transition governing the fragmentation of the call graph plays a key role in the spread of hybrid viruses as well, delimiting the rapid MMS dominated and the slow human mobility driven spreading modes.

In Fig. 4, E and F we explore the additional infective power of a hybrid virus, defined as the ratio $T_{\text{MMS}}(q,m)/T_{\text{H}}(q,m)$ (or $T_{\text{BT}}(q,m)/T_{\text{H}}(q,m)$) relative to its pure MMS (or BT) counterpart. We find that hybrid viruses are about three times faster than an MMS virus at constant market share for $m > m_c$. The contribution of Bluetooth technology for a hybrid virus dominates for $m \leq m_c$, as MMS viruses are unable to spread in this region ($T_{\text{MMS}}(q,m) = \infty$). The additional infective power of a hybrid virus compared to a Bluetooth virus achieves its highest value close to $m_q^*$, decreasing quickly for $m \rightarrow 0$ and mildly for $m \rightarrow 1$, once again underlying the importance of the critical behaviour near $m_q^*$.

Taken together, our results offer a comprehensive picture of the potential dangers posed by mobile viruses. We find that while a Bluetooth virus can reach the full susceptible user base, its spread is slowed by human mobility, offering ample time for developing and deploying countermeasures. In contrast, MMS viruses can reach most susceptible users within hours. Their spread is limited, however, by the market share driven phase transition that fragments the underlying call graph, allowing us to predict that no major virus breakout is expected for OS with market share under the critical point associated with the user base. Therefore, the current lack of major mobile virus outbreak can not be attributed to the absence of effective mobile viruses, but it is mainly rooted in the fragmentation of the call graph. Given, however, the rapid growth in the number of smartphones and the increasing market share of a few OS, it is not unconcievable that the phase transition point will be reached in the near future, raising the possibility of major viral outbreaks. While the most significant danger is posed by hybrid viruses that take advantage of both Bluetooth and MMS protocols, we find that their spread is also limited by the phase transition: hybrid viruses designed for OS with small market share are forced into the slow Bluetooth spreading mode, offering time to develop proper countermeasures. We believe that the understanding of the basic spreading patterns presented here could help estimate the realistic risks carried by mobile viruses and aid the development of proper measures to avoid the costly impact of future outbreaks.

30. We thank D. Brockmann, M. Hypponen, J. Park, Z. Qu, C. Song, A. Vazquez, A. Vespignani and G. Xiao for discussions and comments on the manuscript. We also thank M. Hypponen providing us the valuable mobile phone virus records data. This work was supported by the Defense Threat Reduction Agency Award HDTRA1-08-1-0027, the James S. McDonnell Foundation 21st Century Initiative in Studying Complex Systems, the National Science Foundation within the DDDAS (CNS-0540348), ITR (DMR-0426737) and IIS-0513650 programs and the US Office of Naval Research Award N00014-07-C.

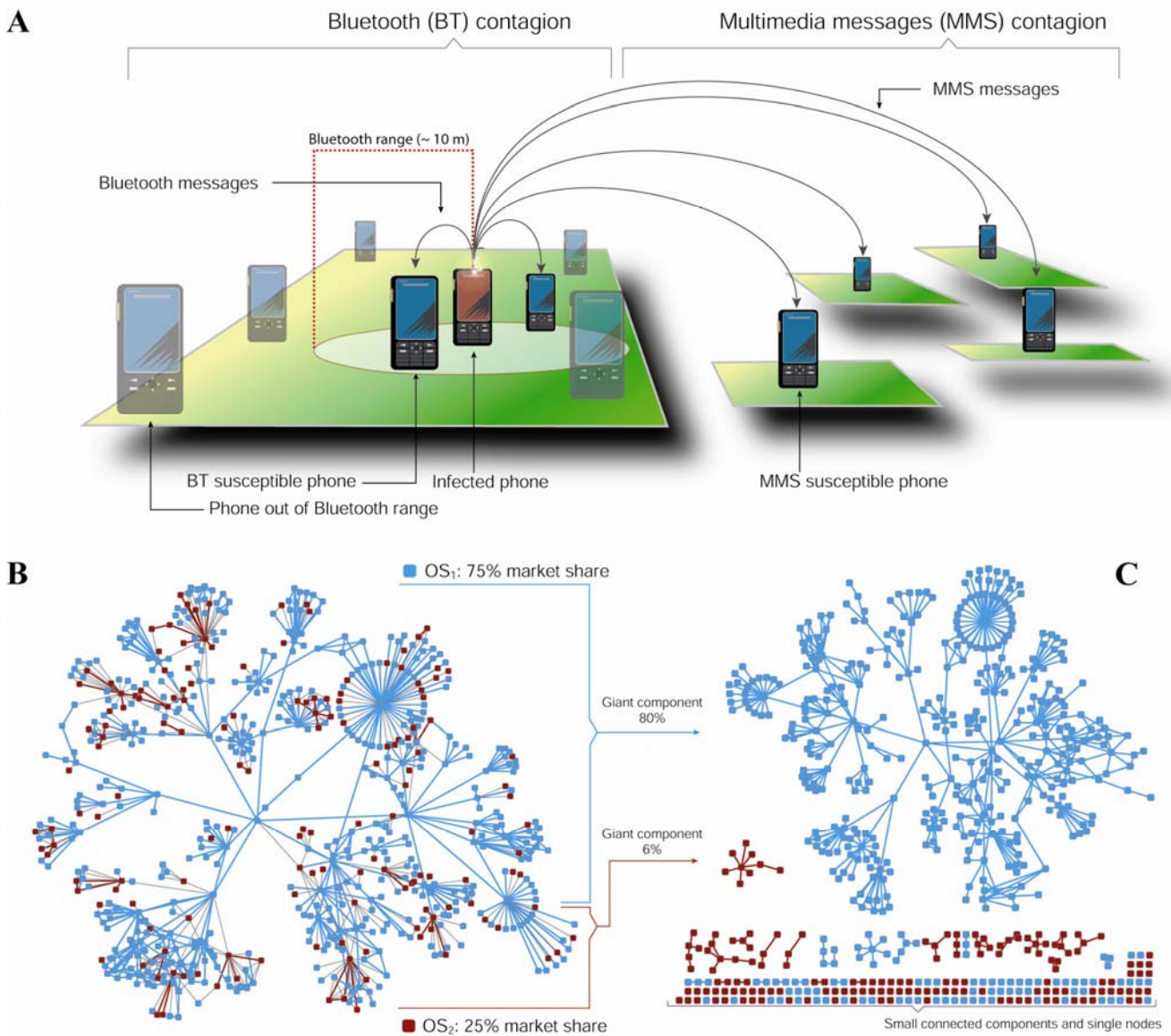

**FIG. 1.** The spreading mechanisms of mobile viruses. (**A**) A Bluetooth virus can infect all phones found within Bluetooth range from the infected phone, its spread being determined by the owner's mobility patterns. An MMS virus can infect all susceptible phones whose number is found in the infected phone's phonebook, resulting in a long-range spreading pattern that is independent of the infected phone's physical location. (**B**) A small neighbourhood of the call graph constructed starting from a randomly chosen user and including all mobile phone contacts up to fourth degrees from it. The color of the node represents the handset's OS, in this example randomly assigned such that 75% of the nodes represent $OS_1$, and the red are the remaining handsets with $OS_2$ (25%). (**C**) The clusters in the call graph on which an MMS virus affecting a given OS can spread, illustrating that an MMS virus can reach at most the number of users that are part of the giant component of the appropriate handset. As the example for the OS shows, the size of the giant component highly depends on the handset's market share (see also Fig. 2C).

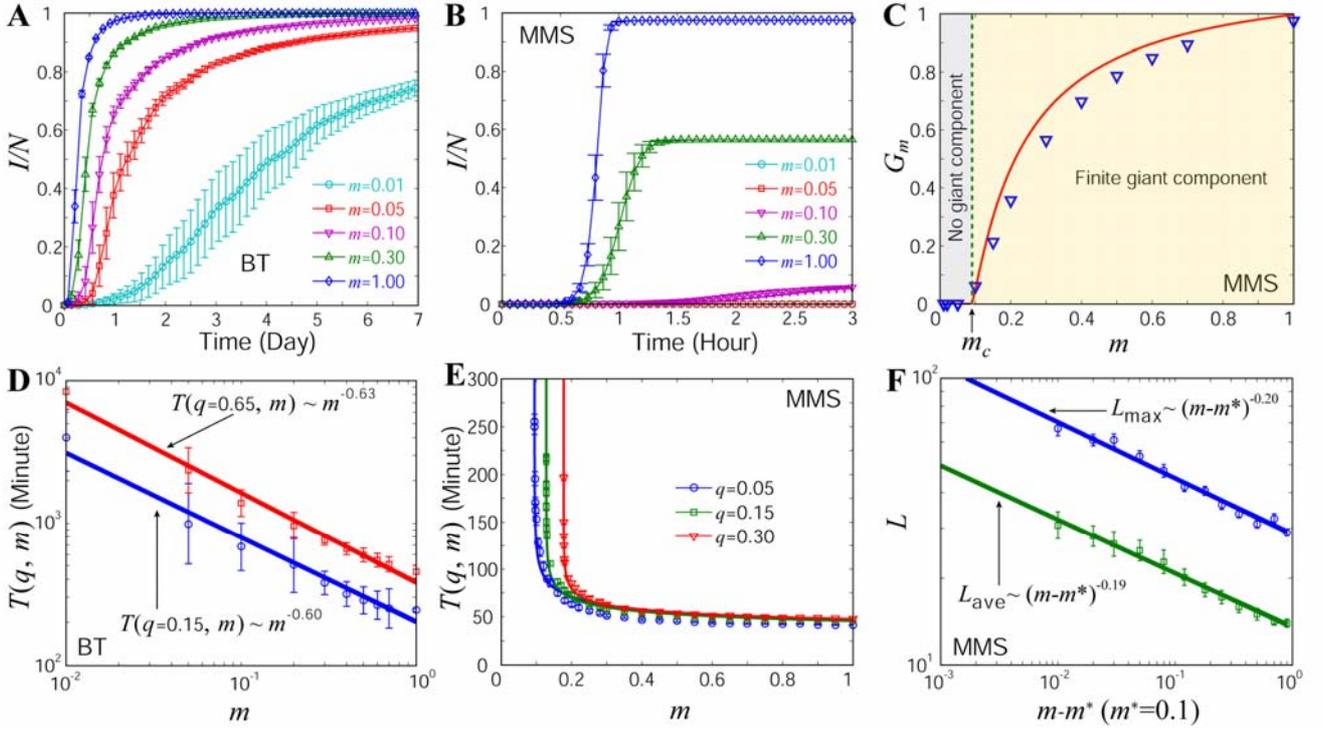

**FIG. 2.** The spreading patterns of Bluetooth and MMS viruses. (**A**) The changes in the ratio of infected and susceptible handsets (*I/N*) with time in the case of a BT virus affecting for handsets with different market share *m*. (**B**) Same as in panel (A), but for MMS viruses, the saturation in *I/N* indicating that an MMS virus can reach only a finite fraction of all susceptible phones. (**C**) The size of the giant component $G_m$ in function of the market share *m*. The blue symbols correspond to the saturation values measured in Fig. 2B, while the red line is the theoretical prediction based on percolation theory (the deviations are attributed mainly to finite size effects and degree correlations, as the calculation assumed an infinite call graph). (**D**) The latency time needed to infect *q*=0.65 or *q*=0.15 fraction of susceptible handsets via a Bluetooth virus, approximated with $T(q=0.65, m) \sim m^{-0.63 \pm 0.05}$ and $T(q=0.15, m) \sim m^{-0.60 \pm 0.04}$ (continuous lines). (**E**) The latency time for an MMS virus for *q*=0.05, 0.15 and 0.30. The continuous lines correspond to $T(q,m) \sim (m - m_q^*)^{-\alpha(q)}$, where the best fits indicate a systematic *q*-dependence, i.e. $\alpha(0.05)=0.20 \pm 0.02$, $\alpha(0.15)=0.17 \pm 0.01$, $\alpha(0.30)=0.14 \pm 0.01$. (**F**) Log-log plot showing the average minimal path length $L_{ave}$ and the longest minimal path length $L_{max}$ for the largest cluster. The fits correspond to $L_{max} \sim (m - m^*)^{-0.20 \pm 0.02}$ and $L_{ave} \sim (m - m^*)^{-0.19 \pm 0.02}$. The curves on (A), (B), (D) are obtained from 10 independent simulations and (E), (F) represent average over 100 runs. For more statistical analysis of the fits in (D), (E), (F), please see the detailed discussion in SOM.

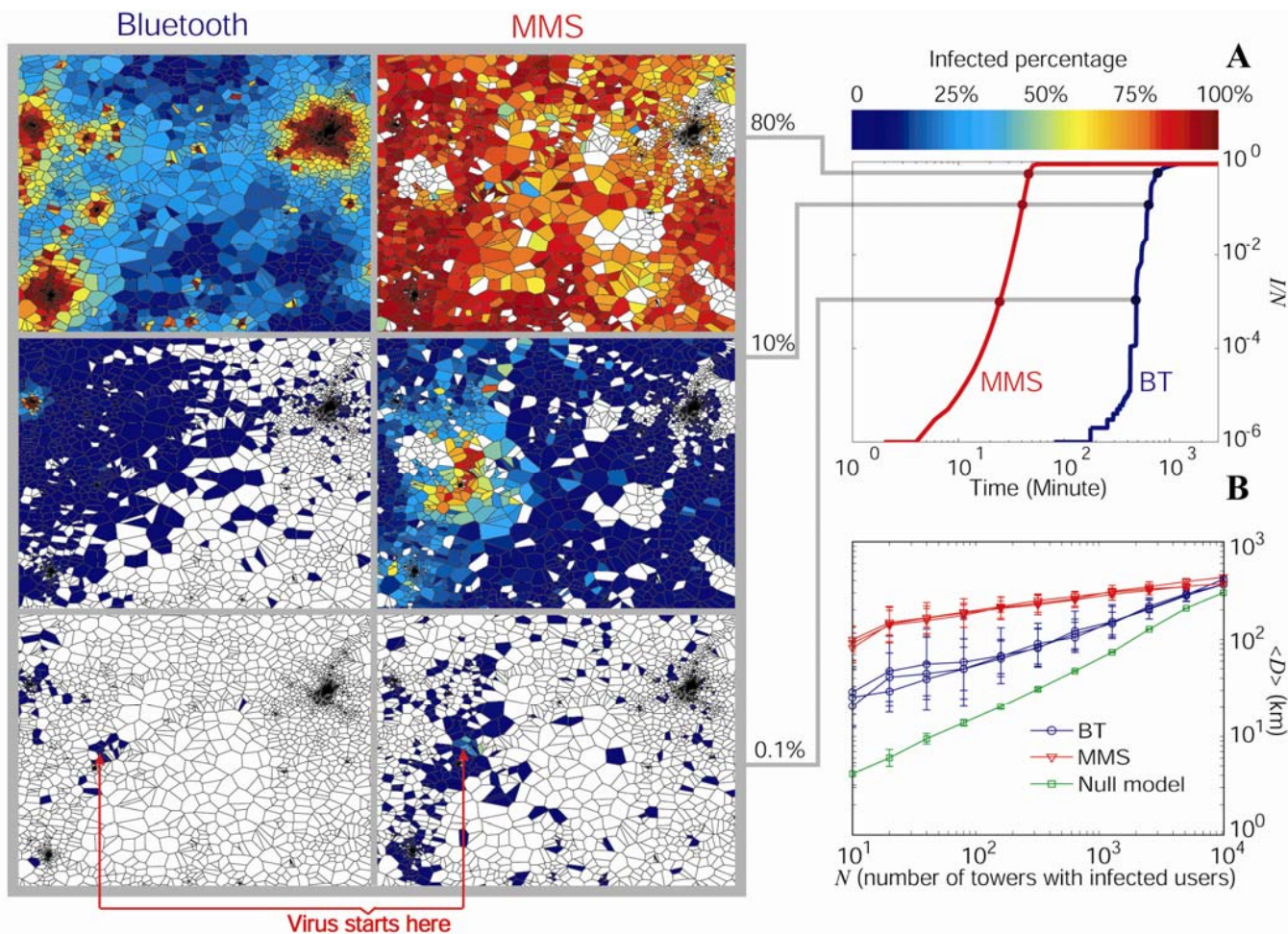

**FIG. 3.** Spatial patterns in the spread of Bluetooth and MMS viruses. (**A**) The virus starts from the same user located at the tower marked by the red arrows. The three panels show the percentage of infected users in the vicinity of each mobile phone tower (denoted by the voronoi cell that approximates each tower's service area). On the right panel we show the corresponding time dependent infection curves, marking the moments when the spatial distribution was recorded. (**B**) Average distance between the tower where the infection was originally started and the most currently infected as a function of $N$ (Three red and blue curves correspond to $m$=0.1, $m$=0.5 and $m$=1), denoting the number of towers with at least one infected users, used as a proxy of time. The green line is obtained from a null model which assumes that the virus can only spread from one tower's service area to its neighbour towers' service areas. The curves on (B) are obtained from 100 independent simulations.

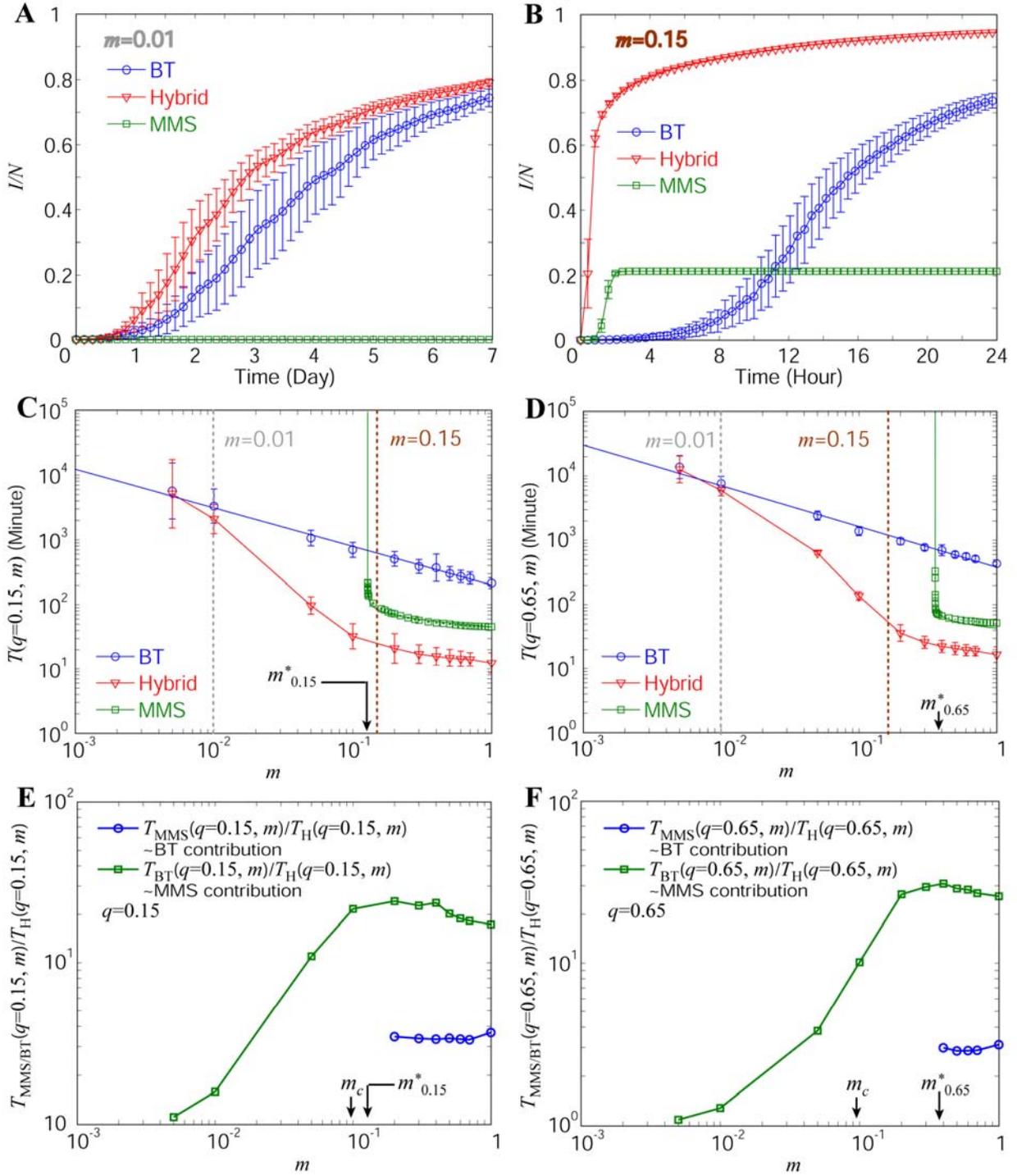

**FIG. 4.** The spreading patterns of hybrid viruses. (**A** and **B**) The time dependent fraction of infected users for a hybrid virus spreading on a handset with (A) *m*=0.01 and (B) *m*=0.15 market share handset, compared with the Bluetooth and MMS spreading modes. (**C** and **D**) The *m*-dependence of latency time for hybrid, MMS and Bluetooth viruses for (C) *q*=0.15 and (D) *q*=0.65. (**E** and **F**) Ratio between the time it takes a Bluetooth or MMS virus to reach (E) 15% and (F) 0.65% of the population divided by the time it takes a hybrid virus to reach the same fraction of the population as a function of the market share *m*. The curves on (A), (B), (C), (D) are obtained from 10 independent simulations.